\documentclass[12pt,oneside,UTF8,AutoFakeBold=4]{article}

\usepackage[top=1in,textheight=8.8in,textwidth=7.1in]{geometry} 
\usepackage{bm,amssymb,amsfonts,amsthm,amscd,mathrsfs,mathtools}
\makeatletter
\renewcommand*\env@matrix[1][\arraystretch]{%
  \edef\arraystretch{#1}%
  \hskip -\arraycolsep
  \let\@ifnextchar\new@ifnextchar
  \array{*\c@MaxMatrixCols c}}
\makeatother
\usepackage{cases}
\usepackage{empheq}
\usepackage{braket} 
\usepackage{enumitem}
\setlist[enumerate,1]{label={(\alph*)}, listparindent=2em}
\usepackage{outlines} 
\usepackage[font=itshape,begintext=``,endtext='']{quoting} 
\usepackage{tabu} 
\usepackage{booktabs} 
\usepackage{graphicx} 
\usepackage{subfigure}
\usepackage{grffile} 
\usepackage[margin=1.1cm, font=footnotesize]{caption} 
\usepackage{float} 
\usepackage{tikz}  
\usetikzlibrary{cd} 
\definecolor{NSFC_blue}{RGB}{0,112,192}
\definecolor{G_red}{RGB}{234,67,53}
\definecolor{G_yellow}{RGB}{251,188,5}
\definecolor{G_green}{RGB}{52,168,83}
\definecolor{G_blue}{RGB}{66,133,244}
\definecolor{G_blue2}{RGB}{0,72,204}
\definecolor{M_red}{RGB}{243,83,37}
\definecolor{M_yellow}{RGB}{255,186,8}
\definecolor{M_green}{RGB}{129,188,6}
\definecolor{M_blue}{RGB}{5,166,240}

\usepackage[backend=biber,style=numeric,sortcites,natbib=true,sorting=none]{biblatex}
\usepackage{algorithm}
\usepackage[noend]{algpseudocode}

\usepackage{nomencl} 
\makenomenclature

\usepackage[unicode,hidelinks,colorlinks=true]{hyperref} 
\usepackage[nameinlink,capitalise]{cleveref} 
\usepackage{titling} 
\thanksmarkseries{fnsymbol}
\usepackage{authblk} 
\newcommand{\rF}{\mathrm{F}}

\newcommand{\bA}{\mathbf{A}}

\newcommand{\bD}{\mathbf{D}}
\newcommand{\bE}{\mathbf{E}}

\newcommand{\bN}{\mathbf{N}}

\newcommand{\bS}{\mathbf{S}}

\newcommand{\bZ}{\mathbf{Z}}

\newcommand{\bmV}{\bm{V}}

\newcommand{\dotx}{\dot{x}}

\newcommand{\bPsi}{\bm{\Psi}}
\newcommand{\bOmega}{\bm{\Omega}}

\theoremstyle{plain}

\theoremstyle{definition}

\theoremstyle{remark}

\newcommand{\ii}{\mathrm{i}}

\newcommand{\T}{\mathsf{T}} 

\newcommand{\p}{\partial}

\makeatletter
\newsavebox{\@brx}
\newcommand{\llangle}[1][]{\savebox{\@brx}{\(\m@th{#1\langle}\)}%
  \mathopen{\copy\@brx\kern-0.5\wd\@brx\usebox{\@brx}}}
\newcommand{\rrangle}[1][]{\savebox{\@brx}{\(\m@th{#1\rangle}\)}%
  \mathclose{\copy\@brx\kern-0.5\wd\@brx\usebox{\@brx}}}
\makeatother

\usepackage{stackengine}
\stackMath
\newcommand\tsup[2][2]{%
 \def\useanchorwidth{T}%
  \ifnum#1>1%
    \stackon[-1.3ex]{\tsup[\numexpr#1-1\relax]{#2}}{\mathaccent"0365{}\kern-.5pt}%
  \else%
    \stackon[-1ex]{#2}{\mathaccent"0365{}\kern-.5pt}%
  \fi%
}

\title{Evaluation of Vortex Criteria by Virtue of the Quadruple Decomposition of Velocity Gradient Tensor
\thanks{This is an English version of the Chinese paper published in Acta Phys. Sin. Vol. 63, No. 5 (2014), doi: 10.7498/aps.63.054704. 
The project was partially supported by the National Basic Research Program of China (973 Program Grant No. 2012CB720101) and the National Natural Science Foundation of China (NSFC Grant No. 11072130).
The translator is the first author (Zhen Li), who is now affiliated with the Beijing Institute of Mathematical Sciences and Applications (BIMSA).}}
\author[1,2]{Zhen Li\thanks{lishen03@gmail.com}}
\author[1]{Xiwen Zhang}
\author[1]{Feng He\thanks{hefeng@tsinghua.edu.cn}}
\affil[1]{School of Aerospace, Tsinghua University, Beijing 100084 China}
\affil[2]{Yau Mathematical Sciences Center, Tsinghua University, Beijing 100084 China}

\begin{document}
\date{(Received 9 October 2013; Revised 27 November 2013)}
\maketitle
\graphicspath{{pics/}}

\abstract
Based on the analysis of the velocity gradient tensor, we investigate in this paper the physical interpretation and limitations of four vortex criteria: $\omega$, $Q$, $\varDelta$ and $\lambda_{ci}$, and reveal the actual physical meaning of vortex patterns which are usually illustrated by level sets of various vortex criteria. A quadruple decomposition based on the normality of the velocity gradient tensor is proposed for the first time, which resolves the motion of a fluid element into dilation, axial stretch along the normal frame, in-plane distortion, and simple shear, in order to clarify the kinematical interpretation of various vortex criteria. The mean rotation characterized by the vorticity $\omega$ always consists of simple shear; the $Q$-criterion can reflect the strength of net rotation within the invariant plane
relative to the axial stretch of a fluid element, and it is a sufficient but unnecessary condition for the existence of net rotation; the $\varDelta$-criterion can exactly identify the existence of net rotation, but it is not the strength of net rotation; in the case that net rotation exists, $\lambda_{ci}$ characterize its absolute strength. net rotation is the total effect of the normal rotation within the invariant plane and the simple shear, where the former is the most basic rotation.
The newly introduced quadruple decomposition can improve our understanding of vortices in fluids and their motions.

\vspace{5mm}

\section{Introduction}

Vortex plays a key role in fluid mechanics. The study of the generation, development, motion, decline of vortices, and the interactions among vortices and those between vortices and solids, and the relationship between vortex and turbulence, as well as the definition and identification of vortices raised in recent years, all have always been important research topics in fluid mechanics.
Shi-Jia Lu and K\"uchemann  \citep{wu_introduction_1993}  call vortex the muscle and essence of fluid motion. \citet{moffatt_kida_ohkitani_1994} referred to vortex as the sinew of turbulence. The analysis of vortices is crucial to understand the motion of fluids. It is a fundamental problem of fluid mechanics.

However, the definition of vortex in fluid mechanics is still vague. The difficulty of defining vortex reflects the distinction between our intuitive understanding and the description methods of classical fluid mechanics. The intuition usually perceives the large-scale change of configuration of fluids, which bears the dynamical-system perspective. However, the theory of classical fluid mechanics usually adopts a field-theoretical description of local properties of the flow fields.

Here we leave aside the justifiability of the definition of vortex and focus on the analysis of existing and applied vortex criteria. They view vortices with a field-theoretical perspective and determine the center of local vortices, which they regard as regions with relatively high vorticity. For any point satisfying a criterion, the fluid element in the neighborhood of the point is regarded as a local vortex rotating about the point. Large vortices are regions consisting of these local vortices.

In a three-dimensional flow field, consider the motion of a point with displacement $(\delta x_1, \delta x_2, \delta x_3)$ relative to its neighbor $O$. The first-order Talyor expansion of its velocity relative to $O$ is
\begin{align}
    \bmV = \bmV_0 + \delta\bmV,
\end{align}
where the relative velocity $\delta\bmV$ is related to the velocity gradient tensor $\nabla\bmV$ as follows:
\begin{align}
    \delta\bmV = (\delta x_1, \delta x_2, \delta x_3)\cdot\nabla\bmV.
\end{align}

The definitions of many vortex criteria are based on the eigen-analysis of the velocity gradient tensor $\nabla\bmV$ of fluid elements, which provides some understanding of the complicated motion of fluid elements. In this paper, a quadruple decomposition based on the normality of the velocity gradient tensor is proposed in an attempt to provide a deeper understanding of the physical meaning and limitations of popular vortex criteria. Then the kinematical properties of local vortices, which are reflected by the quantities of vortex criteria, can be better understood in their applications. It also provides a better understanding of the vortex pattern illustrated by level sets of the quantities. All these are helpful to the analysis and understanding of the patterns of complex flows.

\section{Popular vortex criteria: \texorpdfstring{$\omega$}{ω},\texorpdfstring{$Q$}{Q},\texorpdfstring{$\varDelta$}{∆},\texorpdfstring{$\lambda_{ci}$}{λci},\texorpdfstring{$\lambda_2$}{λ2}}

In experimental or numerical simulation studies of flow fields, people often need to visualize vortical structures. Therefore, various methods are developed for identifying vortices.

For analyzing vortices in flow fields, the simplest way is to use the vorticity field $\omega$, which is referred to as the $\omega$-criterion. $\omega$ is twice the mean angular velocity of the fluid element, hence it reflects the mean rotation. People usually regard the $\omega$-concentrated regions as vortices. The value of $\omega$ is easy to obtain through measurement of velocity profile \citep{Tang_2011}. It is well-known, however, that the vorticity of parallel shear flow is also nonzero, but there are no spiral streamlines in the flow. Hence the criterion has limitations in application.

\citet{okubo_horizontal_1970,hunt_eddies_1988} and \citet{weiss_dynamics_1991} introduced the $Q$-criterion for incompressible or low-compressible flows to determine regions where vorticity tensor exceeds deformation rate tensor. $Q$ is defined as the second of the principal invariants $I_1,I_2,I_3$, i.e., $Q=I_2$. The criterion is $Q>0$, which is widely applied in the identification of vortices in three-dimensional flows \citep{Cai_2011}.

\citet{chong_general_1990} applied the method for analyzing local topological property of equilibrium points of ordinary differential equations to the analysis of complex flow patterns and proposed the $\varDelta$-criterion, where $\varDelta$ is the discriminant of the characteristic polynomial of $\nabla\bmV$. The criterion is $\varDelta>0$. The $\varDelta$ can be expressed in terms of the tree principal invariants of $\nabla\bmV$ as
\begin{align}
\label{eq:Delta_def}
    \varDelta = \frac{1}{4}I_3^2 + \left(\frac{1}{27} I_1^3 - \frac{1}{6}I_1 I_2\right) I_3 + \left(\frac{1}{27}I_2^3 - \frac{1}{108}I_1^2 I_2^2\right).
\end{align}

\begin{figure}[!htbp]
\centering
	\includegraphics[width=0.2\textwidth]{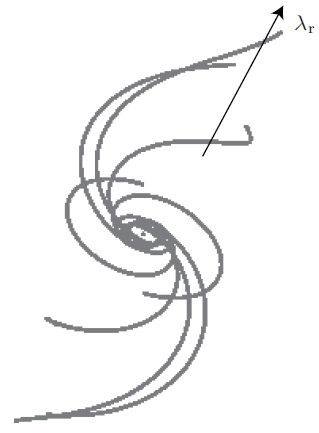}
	\caption{3D helical streamlines in the reference frame moving with the focus.}
	\label{fig:helical_streamlines}
\end{figure}

For any point in the flow field that satisfies $\varDelta>0$, the $\nabla\bmV$ has one real eigenvalue and a pair of complex conjugate eigenvalues. There is a focus of instantaneous streamlines on its invariant plane, which is regarded as the center of a vortex. The instantaneous streamlines around the focus are spiral, which are stretched or compressed along the direction of the eigenvector corresponding to the real eigenvalue, as illustrated in \cref{fig:helical_streamlines}.
The local vortex presented in the invariant plane of $\nabla\bmV$ depends only on the property of eigenvalues of $\nabla\bmV$ at that point. It does not rely on the reference frame of the observer.

\citet{zhou_mechanisms_1999} proposed to use the absolute value of the imaginary part of the complex eigenvalues of $\nabla\bmV$ as the strength of vortex, provided that $\nabla\bmV$ has complex conjugate eigenvalues. It reflects the angular velocity of net rotation within the invariant plane of $\nabla\bmV$ of a fluid element. Since $\lambda_{ci}$ exists only when $\varDelta>0$, it is equivalent to the $\varDelta$-criterion.

Most of the criteria are constructed with the kinematical perspective. But in the research and engineering applications of fluid mechanics, people also use low pressure as an indicator of vortex, for example, this method is widely used for identifying vortices in meteorology. The principle underlying this method is that the rotation of a vortex produces centrifugal force, which must be balanced by the pressure gradient in certain situations. Hence a low-pressure region is created at the center of the vortex. \citet{jeong_identification_1995} modified this method. After some analysis, they concluded that the only reason for the appearance of sectional minimum pressure is the centrifugal force caused by rotation if the effects of viscosity and unsteadiness are excluded. Through some manipulation of the Navier-Stokes equation for incompressible flow, they obtain a tensor $\bD^2+\bOmega^2$, where $\bD$ is the deformation rate tensor and $\bOmega$ is the vorticity tensor. The eigenvalues of this new tensor are arranged in the decreasing order $\lambda_1\ge\lambda_2\ge\lambda_3$. The inequality $\lambda_2<0$ is considered as the necessary and sufficient condition for the pressure to attain its sectional minimum, and it is called the $\lambda_2$-criterion. \citet{chen_new_2011} uses the $\lambda_2$-criterion to illustrate their new findings on the vortices in streamwise steaks of a transient boundary layer.


Many other types of vortex criteria are out of the scope of the current paper. The analysis of the local variables of flow fields belongs to the field theory of classical mechanics, which is easier for computation. Hence the $\omega$-, $Q$-, $\varDelta$-, $\lambda_{ci}$- and $\lambda_2$-criteria are widely applied.

\section{Kinematical analysis of vortex criteria in three-dimensional flow fields}

The second-order tensor $\nabla\bmV$ can be decomposed into a symmetric part and an anti-symmetric part which, expressed in the Cartesian coordinate system, are the deformation rate tensor
\begin{align*}
    D_{ij} = \frac{1}{2}\left(\frac{\p\dotx_j}{\p x_i} + \frac{\p\dotx_i}{\p x_j}\right)
\end{align*}
and the vorticity tensor
\begin{align*}
    \Omega_{ij} = \frac{1}{2}\left(\frac{\p\dotx_j}{\p x_i} - \frac{\p\dotx_i}{\p x_j}\right).
\end{align*}
That is
\begin{align*}
    \nabla\bmV = \bD + \bOmega.
\end{align*}
$\bD$ can be further decomposed as the sum of a dilation tensor $E_{ii}=\p\dotx_i/\p x_i$ and a stretch rate tensor
\footnote{This is a mistake. It should be $E_{ij}=(\nabla\cdot\bmV)/3\delta_{ij}$ and $A_{ij}=D_{ij}-E_{ij}$.}
\begin{align*}
    A_{ij} = \frac{1}{2}\left(\frac{\p\dotx_j}{\p x_i} + \frac{\p\dotx_i}{\p x_j}\right),\quad (i\neq j).
\end{align*}
Hence $\nabla\bmV$ is usually decomposed into three parts (triple decomposition)
\begin{align}
    \nabla\bmV = \bE + \bA + \bOmega.
\end{align}

The motion of a fluid element in a three-dimensional flow is very complex, whose dilation, stretch, and mean rotation are coupled with each other. $\bmV$ is generally non-normal. It is hoped that a new understanding of vortex criteria could be revealed by studying the motion of fluid elements from the perspective of the normality of $\nabla\bmV$.

When $\varDelta\le0$, the eigenvalues of $\nabla\bmV$ $\lambda_1,\lambda_2,\lambda_3$ are all real numbers. $\nabla\bmV$ is represented in the normal frame as an upper triangular matrix
\begin{align}
    \nabla\bmV = 
    \begin{pmatrix}
    \lambda_1 & \gamma & \beta \\
    0 & \lambda_2 & \alpha \\
    0 & 0 & \lambda_3
    \end{pmatrix}.
\end{align}
If the eigenvalues are all real numbers, then there is no focus or center as the topological structure of instantaneous streamlines on the plane spanned by eigenvectors \citep{weiss_dynamics_1991}. Decompose $\nabla\bmV$ as the sum of a normal tensor $\bN$ and a nilpotent tensor $\bS$, where $\bS$ is regarded as the tensor of shear rate in the normal frame.
\begin{align}
    \nabla\bmV = \bN + \bS, \quad
    \bN = 
    \begin{pmatrix}
    \lambda_1 & 0 & 0 \\
    0 & \lambda_2 & 0 \\
    0 & 0 & \lambda_3
    \end{pmatrix}, \quad
    \bS = 
    \begin{pmatrix}
    0 & \gamma & \beta \\
    0 & 0 & \alpha \\
    0 & 0 & 0
    \end{pmatrix}.
\end{align}

The $\bN$ can be further decomposed into a dilation rate tensor $\bE$ representing isotropic dilation, a stretch rate tensor $\bZ$ representing axial stretch along some axis (e.g. the eigenvector associated with $\lambda_3$) and an in-plane tensor $\bPsi$ representing the stretch in the plane orthogonal to the axis. That is
\begin{align}
    \bN = \bE + \bZ + \bPsi,
\end{align}
\begin{align}
\label{eq:normal_tensors_real}
    \bE = \frac{\vartheta}{3}
    \begin{pmatrix}
    1 & 0 & 0 \\
    0 & 1 & 0 \\
    0 & 0 & 1
    \end{pmatrix},\quad
    \bZ = \frac{\varepsilon}{3}
    \begin{pmatrix}
    -1 & 0 & 0 \\
    0 & -1 & 0 \\
    0 & 0 & 2
    \end{pmatrix},\quad
    \bPsi = \psi
    \begin{pmatrix}
    1 & 0 & 0 \\
    0 & -1 & 0 \\
    0 & 0 & 0
    \end{pmatrix},
\end{align}
where $\vartheta = \lambda_1 + \lambda_2 + \lambda_3$, $\varepsilon = \lambda_3 - (\lambda_1 + \lambda_2)/2$, $\psi = (\lambda_1-\lambda_2)/2$.

When $\varDelta>0$, $\nabla\bmV$ has a real eigenvalue $\lambda_r$ and a pair of complex conjugate eigenvalues $\lambda_{cr}\pm \ii \lambda_{ci}$. In the normal frame, there is
\begin{align}
\label{eq:velocity_gradient_normal_frame_complex}
    \nabla\bmV = 
    \begin{pmatrix}
    \lambda_{cr} & \psi + \gamma & \beta \\
    -\psi & \lambda_{cr} & \alpha \\
    0 & 0 & \lambda_r
    \end{pmatrix}.
\end{align}
The characteristic equation of $\nabla\bmV$ is
\begin{align}
    (\lambda - \lambda_r)(\lambda - \lambda_{cr}-\ii\lambda_{ci})(\lambda - \lambda_{cr}+\ii\lambda_{ci}) = 0,
\end{align}
whose discriminant is
\begin{align}
\label{eq:Delta_in_eigenvalues}
    \varDelta = \frac{1}{27}\lambda_{ci}^2\left(\lambda_{ci}^2 + (\lambda_r - \lambda_{cr})^2\right)^2.
\end{align}
Decompose $\nabla\bmV$ as the sum of a tensor $\bN$ and a nilpotent tensor $\bS$, where the expressions of $\bN$ and $\bS$ are as follows:
\begin{align}
    \nabla\bmV = \bN + \bS, \quad
    \bN = 
    \begin{pmatrix}
    \lambda_{cr} & \psi & 0 \\
    -\psi & \lambda_{cr} & 0 \\
    0 & 0 & \lambda_r
    \end{pmatrix}, \quad
    \bS = 
    \begin{pmatrix}
    0 & \gamma & \beta \\
    0 & 0 & \alpha \\
    0 & 0 & 0
    \end{pmatrix}.
\end{align}
And $\bN$ can also be decomposed into three parts
\begin{align}
    \bN = \bE + \bZ + \bPsi,
\end{align}
\begin{align}
\label{eq:normal_tensors_complex}
    \bE = \frac{\vartheta}{3}
    \begin{pmatrix}
    1 & 0 & 0 \\
    0 & 1 & 0 \\
    0 & 0 & 1
    \end{pmatrix},\quad
    \bZ = \frac{\varepsilon}{3}
    \begin{pmatrix}
    -1 & 0 & 0 \\
    0 & -1 & 0 \\
    0 & 0 & 2
    \end{pmatrix},\quad
    \bPsi = \psi
    \begin{pmatrix}
    0 & 1 & 0 \\
    -1 & 0 & 0 \\
    0 & 0 & 0
    \end{pmatrix},
\end{align}
where $\vartheta = \lambda_r + 2\lambda_{cr}$, $\varepsilon = \lambda_r - \lambda_{cr}$.

And the $\psi$ is the absolute value of the imaginary part of the complex conjugate eigenvalues of $\bN$. The physical interpretation of the tensor $\bPsi$ becomes the normal rotation within its invariant plane.

\begin{figure}[!htbp]
\centering
	\includegraphics[width=0.9\textwidth]{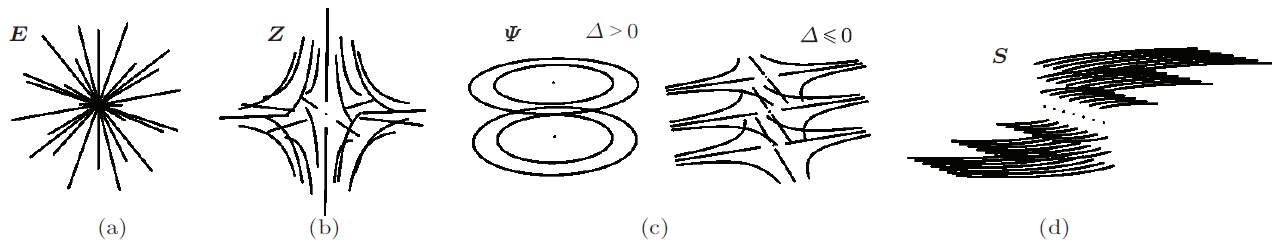}
	\caption{Typical flow patterns corresponding to the tensors in the quadruple decomposition. (a) The dilation rate tensor $\bE$ stands for isotropic expansion and compression; (b) The axial stretch rate tensor $\bZ$ stands for the axial stretch along the direction of frame axis; (c) The in-plane tensor $\bPsi$ orthogonal to the frame axis: $\varDelta>0$, $\bPsi$ stands for the normal rotation in the plane, $\varDelta<0$, $\bPsi$ stands for the stretch in the plane; (d) The shear rate tensor $\bS$ stands for simple shear.}
	\label{fig:typical_flow_patterns}
\end{figure}

For both $\varDelta\le0$ and $\varDelta>0$, $\nabla\bmV$ has the following quadruple decomposition
\begin{align}
\label{eq:4decomposition}
    \nabla\bmV = \bE + \bZ + \bPsi + \bS.
\end{align}
The typical flow patterns corresponding to the tensors $\bE$, $\bZ$, $\bPsi$, $\bS$ are illustrated in \cref{fig:typical_flow_patterns}.

Now we analyze the relationship between the $\bA,\bOmega$ in the usual triple decomposition and the $\bZ,\bPsi,\bS$ in the quadruple decomposition \cref{eq:4decomposition}

Firstly, the $\bS$ is decomposed into a symmetric simple shear $\bS_S$ and an anti-symmetric simple shear $\bS_A$ with
\begin{align}
    \bS_S = \frac{1}{2} (\bS + \bS^\T),\nonumber\\
    \bS_A = \frac{1}{2} (\bS - \bS^\T),
\end{align}
where $\bS^\T$ is the transpose of $\bS$.

When $\bmV$ has three real eigenvalues,
\begin{align}
    \bA =& \bZ + \bPsi + \bS_S,\nonumber\\
    \bOmega =& \bS_A \label{eq:Omega_SA}.
\end{align}

The stretch of the fluid element represented by the tensor $\bA$ consists of three parts: the axial stretch $\bZ$, the in-plane stretch $\bPsi$, and the symmetric simple shear $\bS_S$. The mean rotation represented by $\bOmega$ is just the anti-symmetric simple shear, which is not genuine rotation.

When $\nabla\bmV$ has complex conjugate eigenvalues,
\begin{align}
    \bA =& \bZ + \bS_S, \nonumber\\
    \bOmega =& \bPsi + \bS_A \label{eq:Omega_Psi_SA}.
\end{align}
The stretch of the fluid element represented by the tensor $\bA$ consists of two parts: the axial stretch $\bZ$ and the symmetric simple shear $\bS_S$. The mean rotation represented by $\bOmega$ consists of the normal rotation $\bPsi$ and the anti-symmetric simple shear $\bS_A$. That is to say that the normal rotation is the mean rotation with the part of simple shear $\bS_A$ excluded.

Therefore, no matter in which case of the eigenvalues of $\nabla\bmV$, the stretch rate tensor $\bA$ and vorticity tensor $\bOmega$ consist of parts of the simple shear tensor $\bS$. 

Since the quadruple decomposition separates the simple shear $\bS$, the physical interpretation of the decomposition of the fluid element becomes clear.

According to the triple decomposition of $\nabla\bmV$, the $Q$ introduced by \citet{okubo_horizontal_1970} and \citet{weiss_dynamics_1991} is defined as the second principal invariant $I_2$ in both compressible and incompressible flows,
\begin{align*}
    Q = I_2 = \frac{\vartheta^2}{3} - \frac{\|\bA\|_{\rF}^2}{2} +\frac{\omega}{4}.
\end{align*}
Actually, $I_2$ is affected by the dilation $\vartheta=\nabla\cdot\bmV$ of a fluid element. Even for the compressible flow with uniform dilation, $Q$ is positive, which is obviously unreasonable. Therefore, the relationship between $Q$ and $I_2$ should be (modified to)
\begin{align}
    Q = I_2 - \frac{\vartheta^2}{3}.
\end{align}
Then, for both compressible and incompressible flows, $Q$ has a unified expression
\begin{align}
\label{eq:Q_uniform}
    Q = \frac{1}{4}\left(\omega^2 - 2\|\bA\|_{\rF}^2\right) = \frac{1}{2}\left(\|\bOmega\|_{\rF}^2 - \|\bA\|_{\rF}^2\right).
\end{align}

When $\nabla\bmV$ has three real eigenvalues, plug \cref{eq:Omega_SA} into \cref{eq:Q_uniform}, we get
\begin{align}
    Q = -\left(\frac{1}{3}\varepsilon^2 + \psi^2\right)\le 0.
\end{align}
We can see that $Q$ is a measure of the negative sum of the axial stretch and in-plane stretch of the fluid element.

In the definition \cref{eq:Delta_def} of $\varDelta$, replace the principal invariants with eigenvalues, then considering the relationships among $\vartheta, \psi, \varepsilon$ in \cref{eq:normal_tensors_real}, we can get
\begin{align}
    \varDelta = -\frac{1}{27}\psi^2 (\varepsilon^2 - \psi^2)^2 \le 0.
\end{align}
We can see that $\varDelta$ is a measure of the difference between the axial stretch and in-plane stretch of the fluid element.

When $\nabla\bmV$ has complex conjugate eigenvalues, plug \cref{eq:Omega_Psi_SA} into \cref{eq:Q_uniform}, we get
\begin{align*}
    Q = \psi(\psi + \gamma) - \frac{1}{3}\varepsilon^2,
\end{align*}
or
\begin{align}
\label{eq:Q_lambda_ci}
    Q = \lambda_{ci}^2 - \frac{1}{3}\varepsilon^2.
\end{align}
The \cref{eq:Q_lambda_ci} makes the meaning of $Q$ more significant, which is the difference between the net rotation around the direction of the eigenvector associated with the real eigenvalue $\lambda_r$ and the axial stretch. In the region satisfying $Q>0$, there must be $\lambda_{ci}\neq 0$ ($\delta>0$), i.e., the net rotation exists. 
But in the region of $\lambda_{ci}\neq 0$, the condition $Q>0$ may not be satisfied because the sign of $Q$ is affected by the axial stretch $\varepsilon$. 
Furthermore, \cref{eq:Q_lambda_ci} indicates that the sign of $Q$ is not affected by the existence of simple shear $\bS$.

\begin{figure}[!htbp]
\centering
	\includegraphics[width=0.5\textwidth]{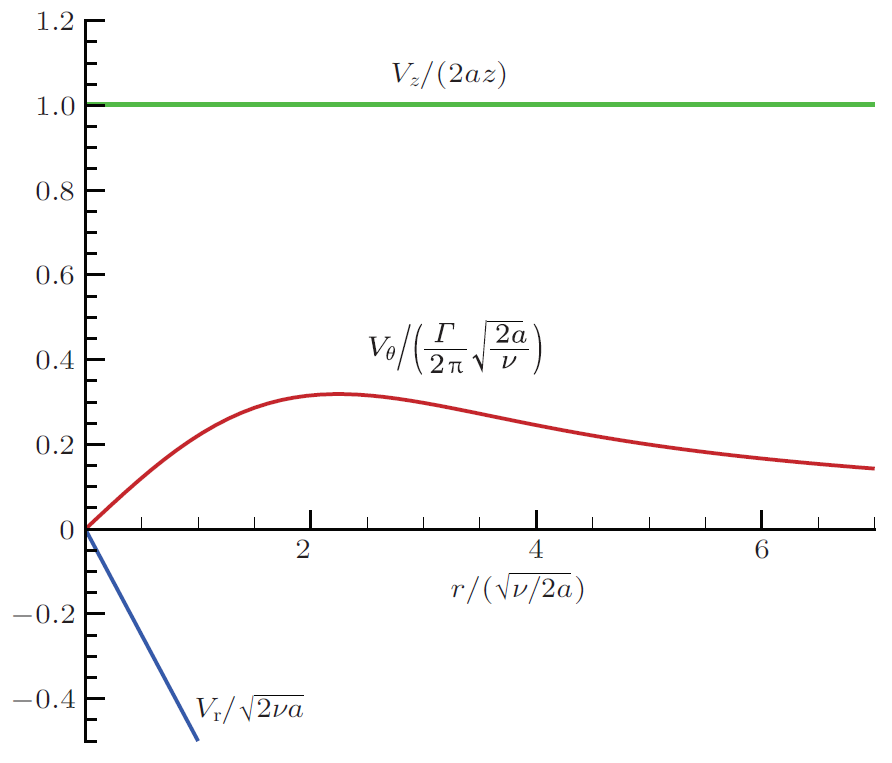}
	\caption{Profiles of velocities of Burgers vortex.}
	\label{fig:Burgers_velocity_profile}
\end{figure}

\begin{figure}[!htbp]
\centering
	\includegraphics[width=0.5\textwidth]{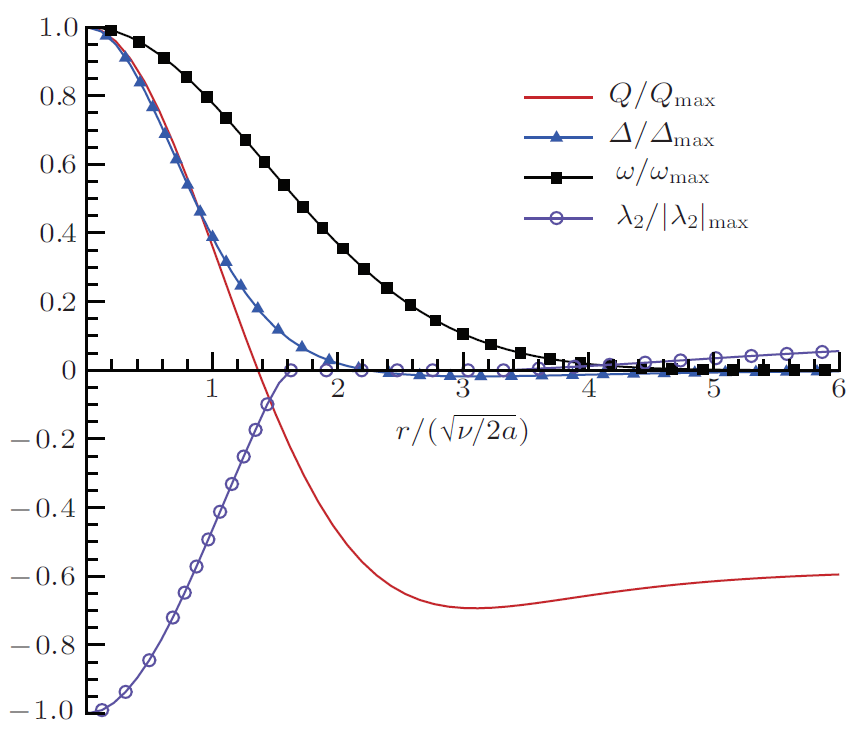}
	\caption{Comparison of size of regions in Burgers vortex determined by the four vortex criteria.}
	\label{fig:Burgers_comparison}
\end{figure}

Let's look at an example of the Burgers vortex, which is axisymmetric and has axial stretching. The radial velocity and axial velocity indicate that the flow has axial stretching. The \cref{fig:Burgers_velocity_profile} shows the profiles of the axial velocity $V_z$, the radial velocity $V_r$ and the azimuth velocity $V_{\theta}$ of a Burgers vortex with strong axial stretching. The \cref{fig:Burgers_comparison} shows the comparison of the four vortex criteria on this flow.

\begin{align*}
    0.00 < r/\sqrt{\frac{\nu}{2a}} < 1.38,&\quad
    Q > 0,\quad
    \Delta > 0,\quad
    \lambda_{ci}\neq0,\quad
    \omega\neq0,\quad
    \lambda_2 < 0;\\
    1.38 < r/\sqrt{\frac{\nu}{2a}} < 1.63,&\quad
    Q < 0,\quad
    \Delta > 0,\quad
    \lambda_{ci}\neq0,\quad
    \omega\neq0,\quad
    \lambda_2 < 0;\\
    1.63 < r/\sqrt{\frac{\nu}{2a}} < 2.20,&\quad
    Q < 0,\quad
    \Delta > 0,\quad
    \lambda_{ci}\neq0,\quad
    \omega\neq0,\quad
    \lambda_2 \ge 0;\\
    2.20 < r/\sqrt{\frac{\nu}{2a}} < 6.00,&\quad
    Q < 0,\quad
    \Delta < 0,\quad
    \lambda_{ci}\, \text{doesn't exist},\quad
    \omega\neq0,\quad
    \lambda_2 \ge 0.
\end{align*}

When $\nabla\bmV$ has complex conjugate eigenvalues, plug the $\vartheta,\varepsilon$ in \cref{eq:normal_tensors_complex} and their relationships with eigenvalues into \cref{eq:Delta_in_eigenvalues}, we get
\begin{align}
\label{eq:Delta_lambda_ci}
    \varDelta = \frac{1}{27}\lambda_{ci}^2 \left(\varepsilon^2 + \lambda_{ci}^2\right)^2 > 0.
\end{align}

The \cref{eq:Delta_lambda_ci} makes the physical interpretation of $\varDelta$ more significant too. It measures the sum of the net rotation about the eigenvector associated with the real eigenvalue $\lambda_r$ and the axial stretch, where the axial stretch $\varepsilon$ does not affect the validity of $\varDelta>0$. Hence the $\varDelta$-criterion is suitable for determining the qualitative structure of local streamlines. However, as a measure of the absolute strength of net rotation, it is affected by the axial stretch.

The discriminant obtained from \cref{eq:velocity_gradient_normal_frame_complex} should be identical to \cref{eq:Delta_in_eigenvalues}, which gives 
\begin{align}
\label{eq:geometric_mean}
    \lambda_{ci}^2 = \psi (\psi + \gamma).
\end{align}
The existence of $\lambda_{ci}$ indicates most clearly that there is net rotation within the invariant plane of a fluid element. Its absolute value or square can be used as a measure of the absolute strength of net rotation. From \cref{eq:geometric_mean} we can see that the net rotation represented by $\lambda_{ci}$ is a total effect of the normal rotation $\psi$ within its invariant plane and the simple shear $\gamma$.

According to \cref{eq:Omega_Psi_SA}, the normal rotation $\psi$ is the mean rotation with part of simple rotation removed, and it is the most basic rotation. It is contained in the four vortex criteria that appeared in this paper. In theory, it can be used as the most basic local vortex criterion. But its computation is a little more complicated than the four popular vortex criteria.

Although the $\lambda_2$-criterion is motivated by dynamical considerations, its definition is completely a kinematical quantity. The kinematical meaning of the $\lambda_2$-criterion is far less significant than the $\varDelta$- and $\lambda_{ci}$-criteria. The second principal invariant of $\bD^2+\bOmega^2$ consists of the coupled dilation, stretch, and mean rotation of a fluid element. Hence the phenomenon described by $\lambda_2$ is completely three-dimensional, whose complexity prevents a decomposition into several parts like the other four vortex criteria. It is a compound of all parts. A deeper discussion about the $\lambda_2$-criterion will be left to future study.
\section{Conclusions}

By decomposing the velocity gradient tensor $\nabla\bmV$ into the isotropic dilation rate tensor $\bE$, the axial stretch tensor $\bZ$, the in-plane tensor $\bPsi$ and the simple shear tensor $\bS$, we can see that the mean rotation represented by the vorticity of the fluid element always consists of components of simple shear $\bS$. Therefore, the condition of nonzero vorticity can not distinguish parallel shear flow. The value of $Q>0$ represents the strength of net rotation relative to axial stretch, but it is only a sufficient and non-necessary condition for the existence of net rotation within the invariant plane of $\nabla\bmV$. The sign of $\varDelta$ determines directly the meaning of in-plane tensor $\bPsi$: whether it is in-plane stretch or normal rotation in the invariant plane of $\bPsi$. Also, it can exactly identify the existence of net rotation within the invariant plane of $\nabla\bmV$. However, its value is not the strength of net rotation due to the influence of axial stretch. The $\lambda_{ci}$ is a measure of the absolute strength of net rotation within the invariant plane of $\nabla\bmV$ of a fluid element, which is a combination of normal rotation within its invariant plane and simple shear. Normal rotation is the mean rotation with part of the simple shear removed. It is a basic rotation, but its computation is relatively complicated as a vortex criterion. The sizes of regions determined by the four vortex criteria are generally different. Their values reflect more about different kinematical features of the determined regions. The meanings of vortex patterns illustrated by level sets of different vortex criteria are also different. The $\lambda_{ci}$-criterion is relatively simple. The $\lambda_2$-criterion is initiated based on an analysis of dynamics, but its definition is a kinematical quantity. Due to its complex characteristics, it can not be decomposed into several parts in a simple way for analysis.

\printbibliography
\end{document}